\documentclass[pasp]{emulateapj}

\newcommand{\pa}{Pa$\alpha$}
\newcommand{\pb}{Pa$\beta$}
\newcommand{\ha}{H$\alpha$}
\newcommand{\lya}{Ly$\alpha$}
\newcommand{\kms}{{\rm km~s\ensuremath{^{-1}}}}

\usepackage{natbib}
\usepackage{multirow}
\usepackage{appendix}
\usepackage{longtable}

\slugcomment{PASP Accepted}

\shorttitle{The Future of Direct SMBH Masses}
\shortauthors{Batcheldor \& Koekemoer}

\begin{document}


\title{The Future of Direct Supermassive Black Holes Mass Estimates}


\author{D. Batcheldor}
\affil{Center for Imaging Science, Rochester  Institute of Technology}
\affil{54 Lomb Memorial Drive, Rochester, NY, 14623, USA}
\email{dan@astro.rit.edu}
\and
\author{A. M. Koekemoer}
\affil{Space Telescope Science Institute}
\affil{3700 San Martin Drive, Baltimore, MD, 21218, USA}
\email{koekemoer@stsci.edu}



\begin{abstract}
The repeated discovery of supermassive black holes (SMBHs) at the centers of galactic bulges, and the discovery of relations between the SMBH mass ($M_\bullet$) 
and the properties of these bulges, has been fundamental in directing our understanding of both galaxy and SMBH formation and evolution. However, 
there are still many underlying questions surrounding the SMBH - galaxy relations. For example, are the scaling relations linear and constant throughout cosmic history, 
and do all SMBHs lie on the scaling relations? These fundamental questions can only be answered by further high quality direct $M_\bullet$ estimates from a wide range in 
redshift, before further refinements to galaxy evolution models can be made. In this paper we determine the observational requirements necessary to directly determine 
SMBH masses, across cosmological distances, using current $M_\bullet$ modeling techniques. We also discuss the SMBH detection abilities of future facilities. We 
find that if different $M_\bullet$ modeling techniques, using different spectral features, can be shown to be consistent, then both 30~m ground- and 16~m space-based 
telescopes will theoretically be able to sample $M_\bullet\sim10^{9}M_\odot$ across $\sim95\%$ of cosmic history. In addition, SMBHs as small as $10^{6}M_\odot$ 
will be sampled at a distance of the Coma cluster, and  SMBHs as small as $10^{4}M_\odot$ will be sampled in the Local Group. However, we find that the abilities of 
ground-based telescopes critically depend on future advancements in adaptive optics systems; more limited AO systems will result in limited effective spatial resolutions, 
i.e., SMBH detection efficiency, and forces observations towards the near-infrared where spectral features are weaker and more susceptible to sky features. Ground-based 
AO systems will always be constrained by relatively bright sky backgrounds and atmospheric transmission. The latter forces the use of multiple spectral features and 
dramatically impacts the SMBH detection efficiency. The most efficient way to advance our database of direct SMBH masses is therefore through the use of a large (16~m) 
space-based UVOIR telescope.
\end{abstract}




\section{Introduction}

Direct mass estimates of supermassive black holes (SMBHs) are made by spatially resolving the gravitational influence of the SMBH 
itself. There is a constantly growing database of such direct SMBH mass ($M_\bullet$) estimates (e.g. \citealt{2008PASA...25..167G,2009ApJ...698..198G}) 
that has been repeatedly used to demonstrate intimate links between $M_\bullet$ and rudimentary properties of the surrounding host galaxy.  For example, 
$M_\bullet$ is seen to scale with the bulge luminosity \citep{1995ARA&A..33..581K}, the stellar velocity dispersion,  $\sigma_\ast$ 
\citep{2000ApJ...539L...9F,2000ApJ...539L..13G}, the bulge concentration index \citep{2001ApJ...563L..11G},  the stellar light deficit 
\citep{2009ApJ...691L.142K}, and potentially the dark matter halo mass \citep{2002ApJ...578...90F,2003MNRAS.341L..44B,2005ApJ...631..785P}.  

These SMBH scaling relations have had a fundamental impact on our understanding of galaxy formation and evolution. For example, it is possible 
to use scaling relations to non-directly estimate and extrapolate  $M_\bullet$ from galactic bulges in which the central SMBH's gravitational 
influence is not resolved, i.e., distant and/or small bulges.  Therefore, scaling relations could be used to estimate $M_\bullet$ across cosmic history 
and constrain the black hole mass function (BHMF). Consequently, scaling relations have fostered a wealth of exciting new theoretical 
investigations \citep{2001ApJ...552L..13C,2003ApJ...591..125A,2005MNRAS.364..407C} and have potentially placed important limits to 
evolutionary models \citep{2004ApJ...613..109H,2005ApJ...634..910W}. 

Due to its small scatter the $M_\bullet-\sigma_\ast$ relation has received the most attention. However, \cite{2003ApJ...589L..21M} and 
\cite{2007MNRAS.379..711G} have shown that with careful morphological and multi-wavelength analyses, scaling relation scatters can be 
comparable. The $M_\bullet-\sigma_\ast$ relation, characterized as $\log{M_\bullet}=\alpha+\beta\log{\sigma_\ast}$, has had many attempts 
to fit its zero-point ($\alpha$), slope ($\beta$) and scatter \citep{2000ApJ...539L...9F,2000ApJ...539L..13G,2001ApJ...547..140M,2002ApJ...574..740T}. 
Unfortunately, these individual fits have produced considerably different results in which $\beta$ ranges from 4.0 to 4.9, for example. In addition, both 
the intermediate mass black hole (IMBH,  $M_\bullet<10^5M_\odot$) and hyper-massive black hole (HMBH,  $M_\bullet>10^{10}M_\odot$) 
regimes of the BHMF are almost entirely unexplored. This leaves the linearity of the $M_\bullet-\sigma_\ast$ relation unclear. Consequently, 
we do not know the upper and lower limits to the BHMF, or even if such limits exist. 

If we assume that all SMBHs lie on the $M_\bullet-\sigma_\ast$ relation, that the relation is linear, and that fainter less massive galaxies contain 
SMBHs rather than compact stellar nuclei \citep{2006ApJ...644L..21F}, then taking $\sigma_\ast=20\kms$, a value found in large globular clusters 
\citep{1995A&A...303..761M} and where IMBHs may reside, the $\beta=4.0$ slope predicts $\log{M_\bullet}= 4.1(+0.4,-0.4) M_\odot$ while the 
$\beta=4.9$ slope predicts $\log{M_\bullet}=3.4(+0.4,-0.5) M_\odot$. Taking $\sigma_\ast=444\kms$, a value seen in brightest cluster galaxies 
\citep{2008ApJ...687..828S} and where HMBHs may reside, the $\beta=4.0$ slope predicts $\log{M_\bullet}=9.5(+0.2,-0.1) M_\odot$ while the 
$\beta=4.9$ slope predicts $\log{M_\bullet}=9.9(+0.2,-0.2) M_\odot$. Therefore, high accuracy direct estimates of $M_\bullet$ from IMBHs and 
HMBHs are of critical importance. A poorly determined $M_\bullet-\sigma_\ast$ relation introduces large uncertainties in extrapolated values of 
$M_\bullet$, and influences our understanding of both SMBH and galaxy formation and evolution. 

However, it must be noted that scaling relations established in the local universe may have experienced cosmic evolution 
\citep[][and references therein]{2007ApJ...667..117T}. Therefore, if evolutionary models are to be properly constrained, it must be determined whether the 
scaling relationships themselves evolve. Consequently, an accurate and complete sample of SMBH masses must be constructed 
from as wide a range of redshifts as possible. Such a database would then shift the importance of the scaling relations, as extrapolation 
of the scaling relations would no longer be required to constrain evolutionary models.

The SMBHs that power QSOs frequently have their masses estimated using methods, such as reverberation mapping 
\citep{2004ApJ...613..682P}, that are calibrated from the scaling relations based on the unknown geometry of the inner broad line region (BLR). 
The enormous luminosities of QSOs make them ideal candidates for establishing SMBH masses at high redshift. However, without well understood 
scaling relations the reverberation mapping calibration remains uncertain. Direct estimates if $M_\bullet$ in QSOs, while challenging, will determine 
if QSOs follow the same scaling relations as established in quiescent galaxies, and will provide an absolute calibration for continued reverberation 
mapping of high redshift SMBHs. In the closest (un-obscured) QSO, 3C273, assuming $M_\bullet = 10^{9.8}M_\odot$ \citep{2005A&A...435..811P} 
at a distance of 640 Mpc, a minimum resolution of 0\farcs05 ($\sim150$pc) is needed to directly determine $M_\bullet$. However, using current 
techniques it is also a challenge to determine an accurate value of $\sigma_\ast$, as the QSO can drown out the signal from the host galaxy
and fill in the absorption features needed to determine $\sigma_\ast$.

The future path of SMBH investigations is likely to be determined by the questions posed above. These can be summarized as follows:

\begin{itemize}
\item{Do HMBHs follow the same scaling relations as defined by SMBHs?}
\item{Do IMBHs or compact stellar nuclei exist in fainter less massive galaxies?}
\item{Are SMBH scaling relations linear?}
\item{Have SMBH scaling relations evolved through cosmic history?}
\item{What are the upper and lower limits to the BHMF?}
\item{What are direct SMBH masses in QSOs and do they follow the established scaling relations?}
\end{itemize}

In this paper we present the observational requirements necessary to answers these questions. In \S~\ref{estimate}  we discuss the 
current modeling techniques that can be applied to large samples of galactic bulges in order to make direct $M_\bullet$ estimates. In \S~\ref{ifs} we 
discuss the value of integral field spectroscopy in accurate $M_\bullet$ estimates, and in \S~\ref{rh} we determine the size of telescope required 
to resolve the gravitational influence of a SMBH. \S~\ref{sense} discusses the potential sensitivities of both ground- and space-based facilities. In 
\S~\ref{future} we discuss the SMBH detection abilities of up-coming and proposed telescopes, and \S~\ref{conclusions} sums up. 

\section{Direct Models of $M_\bullet$}\label{estimate}

There are a number of direct $M_\bullet$ estimate techniques, many of which are only applicable in special cases. For example, due to the extremely high 
spatial resolution required, proper motion studies can only be carried out around SagA*, and H$_2$0 Masers can only be used if the plane of the rotating 
gas is aligned very close to our line-of-sight. However, at present, there are two methods that can be applied to significant galaxy populations. Models 
of the nuclear gas kinematics (e.g., \citealt{1996ApJ...470..444F,1997ApJ...489..579M,2001ApJ...549..915M}) and stellar dynamics (e.g., 
\citealt{1994MNRAS.270..271V,2002MNRAS.335..517V,2003ApJ...583...92G}). 

Models of gas kinematics are conceptually straight-forward (a rotating Keplerian disk) provided that a nuclear gas disk is in-fact present and has a well 
defined inclination. In addition, the disk must be dominated by the gravitational influence of the SMBH, rather than by any inflows, outflows or 
turbulences. The narrow \ha\ and [NII] emission lines are typically used to build a nuclear rotation curve, however, it remains unclear as to whether 
other strong emission lines from other species, e.g., CIV, CIII], MgII, [OIII], HeI, produce consistent $M_\bullet$ estimates. For example, observed narrow 
emission lines are similar to the emission lines observed in planetary nebulae and H II regions, both of which experience strong non-gravitational kinematics. 
A final complication is introduced if there is an active galactic nucleus (AGN) present. AGN produce spatially unresolved broad emission lines that must be 
subtracted in order to construct a clean rotation curve. However, the use of strong emission lines means that high signal to noise (S/N) observations can be made in 
a relatively short time. 

Stellar dynamical models do not suffer from many of the draw backs of gas kinematics (possible non-presence of a nuclear disk, unconstrained disk inclinations, 
non-gravitational motions) but it is also unclear as to whether different absorption features, e.g., Ca H \& K, Mgb, CaT, CO band-heads, produce 
consistent $M_\bullet$ estimates. The use of stellar absorption features means that high continuum S/N is required in order to fit the line-of-sight 
velocity distributions (LOSVDs) . The low surface brightness of many bulges means that the required S/N can be difficult to achieve using a practical 
exposure time. Finally, stellar dynamical models are notoriously complex and require a large amount of time to completely explore $\chi^2$ space. 
In addition, there is a possible degeneracy in the stellar dynamical models that can produce a significant range of consistent SMBH masses 
\citep{2004ApJ...602...66V}.

Despite their respective disadvantages, gas and stellar dynamical models currently offer the only methods that can produce significant populations of 
homogeneous $M_\bullet$ estimates. However, it must be noted that while almost 90\% of current direct estimates have been made using these methods 
\citep{2008PASA...25..167G}, it is still unclear as to whether the two methods in themselves produce consistent results. Nevertheless, gas and stellar 
dynamical studies have shown three requirements of data capable of directly determining $M_\bullet$ in a galactic bulges. These requirements are 
explained in more detail in the proceeding sections.  First, integral field spectroscopy (IFS) must be used to accurately determine the stellar contribution 
to the total gravitational potential. Second, the stellar potential needs to be separated from the contribution of the SMBH by resolving the sphere of influence 
radius ($r_h$). Finally, the S/N of the data (instrument sensitivity) must be high enough to accurately fit line profiles and LOSVDs.

\section{Integral Field Spectroscopy}\label{ifs}

IFS is used to determine the stellar gravitational potential and provide the large number of data points essential for fully exploring $\chi^2$ space in 
dynamical models \citep{2004ApJ...602...66V}. However, the value of IFS stretches beyond its use in dynamical models to defining the properties of 
the bulge itself. A problem with the fitting of the $M_\bullet-\sigma_\ast$ relation concerns the values of $\sigma_\ast$ used 
\citep{2000ApJ...539L...9F,2000ApJ...539L..13G}. The size and shape of the spectroscopic aperture used to observe the LOSVD has an impact on the 
fitted parameters \citep{2005ApJS..160...76B}. Therefore, IFS is an essential tool in estimating a robust $M_\bullet-\sigma_\ast$ relation as any 
2D aperture can be used to determine $\sigma_\ast$ after the data has been collected.  

An integral field unit (IFU) optimized to detect SMBHs will be able to spatially over-sample $r_h$. In addition, the 
IFU must have a field of view able to spatially sample out to an effective radius (typically  $\sim1$kpc, \citealt{2003ApJ...589L..21M}) to accurately 
define the gravitational potential of the host bulge. The spectral pixel size of the IFU could vary across the field of view. 
Maximum spatial resolution is required at the center of the array, while spatial sampling over a larger area can be used at greater radii in order to 
collect enough S/N from lower surface brightness regions. Finally, the IFU must be able to sample both ends of the BHMF, e.g., IMBHs in globular 
clusters and dwarf spheroidals ($\sigma_\ast\approx10\kms$), and HMBHs in QSOs. Therefore, it must have high spectral resolution 
($\Re=\lambda/\delta\lambda\sim30,000$) and be able to detect the stellar continuum of QSO hosts while not saturating the detector at the QSO 
nucleus. 

Emissions from QSOs dominate those from the host galaxy. However, an IFU could directly exclude contributions from active nuclei through the use of 
a micro mirror array. Alternatively, the active nucleus and host could simply be observed using an extreme dynamic range array such as the Reticon 
\citep{1990ASPC....8..100C}. These detectors have full wells capable of holding $10^9$ electrons before affecting adjacent pixels. Charge injection 
devices, such as SpectraCam \citep{2005SPIE.5902...73B}, may also be able to simultaneously observe both bright and faint sources in the same field of 
view, without compromising the integrity of the detector, or the science image, as individual pixels are read out when the limit of the full well is 
approached. In addition to observations of active galaxies, such detectors will also be valuable asset in the study of many other important astrophysical 
objects, and properties, that require a high dynamic range, e.g., the low end of the stellar initial mass function, supernova ejecta, stellar debris disk and 
extra-solar planets. 

\begin{figure}
\plotone{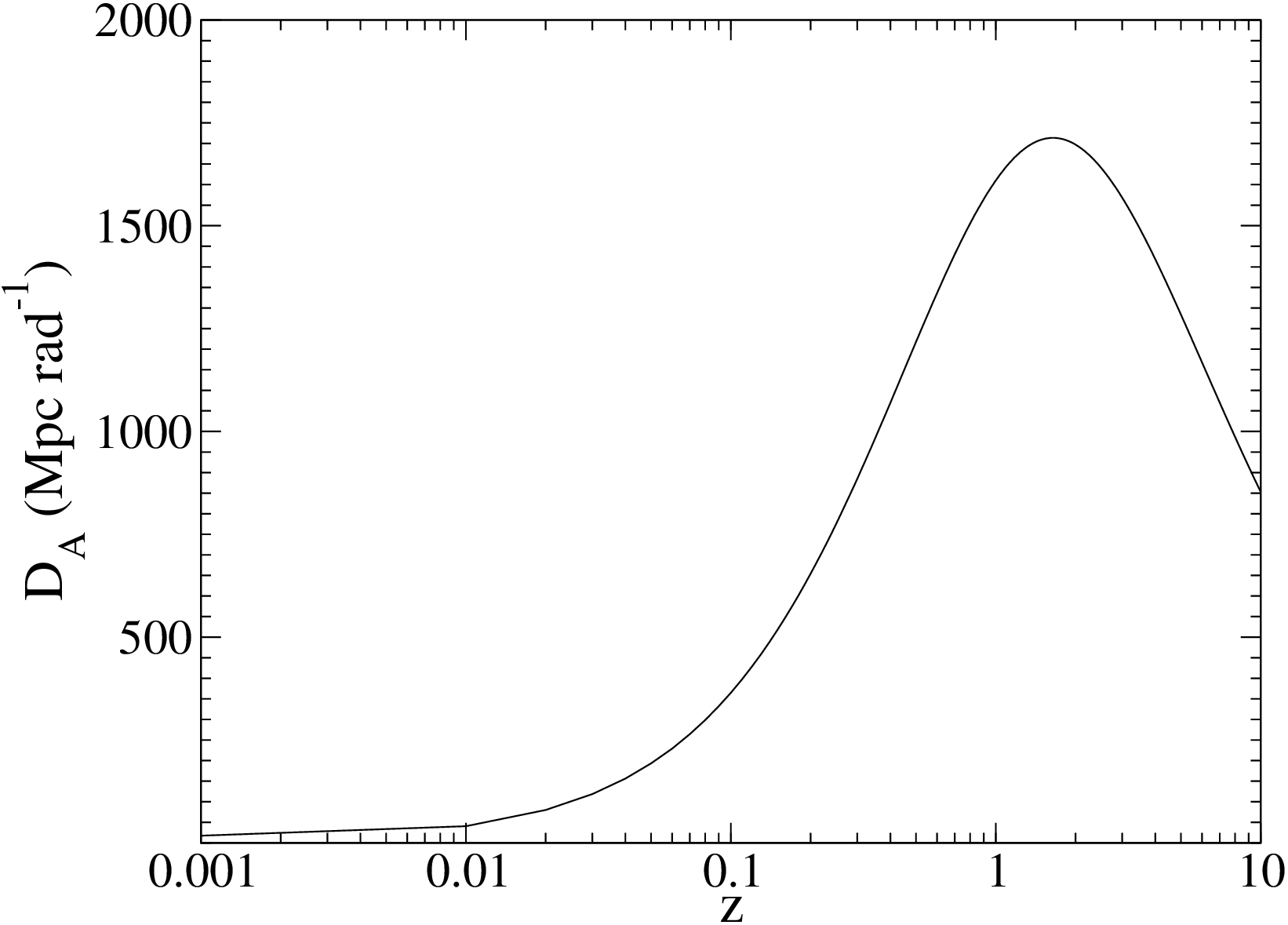}
\caption{Angular diameter distance ($D_A$) as a function of redshift assuming the standard cosmology 
($\Omega_m=0.27, \Omega_\lambda=0.73$ and $H_0=73\kms{\rm~Mpc^{-1}}$). 
}
\label{fig:1}
\end{figure}

\section{Required Spatial Resolution}\label{rh}

In considering a collapsed object at the center of a distribution of stars, \cite{1972ApJ...178..371P} first derived a characteristic radius given 
by Equation~\ref{equ:rh}. This radius, which has since become known as the ``sphere of influence'', has been used to estimate the spatial 
scale at which the potential of the SMBH dominates over that of the host galaxy. 

\begin{equation}\label{equ:rh}
r_h = \frac{GM_\bullet}{\sigma_\ast^2}
\end{equation}

Spatially resolving $r_h$ is considered to be a requirement for detecting the dynamical signature of a SMBH \citep{2002ApJ...578...90F,2003ApJ...589L..21M,2004ApJ...602...66V}, 
and therefore directly measuring $M_\bullet$. It then follows that the $M_\bullet-\sigma_\ast$ relation can be used to determine the ability of a telescope to resolve 
$r_h$ given its diffraction limit ($\theta_D$). However, to investigate cosmic evolution of the SMBH scaling relations, $M_\bullet$ must be determined across a large 
range of redshift, i.e., cosmological affects must be taken into consideration.

When considering spatial resolution as a function of redshift, the cosmological angular diameter distance ($D_A$) is of fundamental importance. 
It is the ratio of physical size to the angle subtended on the sky, i.e., the SMBH sphere of influence radius to the diffraction limit of the 
telescope. Therefore, in determining $D_A$ we can calculate the relation between the angular size of $r_h$ and redshift, and subsequently the 
ability of a diffraction limited telescope to make direct determinations of $M_\bullet$. 

The tangential co-moving distance ($D_M$) is simply related to $D_A$ by Equation~\ref{equ:datodm} (where $z$ is redshift). 

\begin{equation}\label{equ:datodm}
D_A = \frac{D_M}{(1+z)}
\end{equation}

In a flat Universe ($\Omega_k=0.00$) $D_M$ is equal to $D_C$ (the radial co-moving distance). Therefore, following \cite{1993ppc..book.....P}
it can be shown that $D_A$ is given by Equation~\ref{equ:ztoda}.

\begin{equation}\label{equ:ztoda}
D_A = \frac{c}{H_0(1+z)}\int^z_0 (\Omega_M(1+z')^3 + \Omega_\lambda)^{-\frac{1}{2}}dz'
\end{equation}

Assuming the standard cosmological model, with $\Omega_M=0.27, \Omega_\lambda=0.73$ and $H_0=73\kms{\rm~Mpc^{-1}}$, we can see in Figure~\ref{fig:1} 
how $D_A$ varies as a function of redshift. This demonstrates a turnover in $D_A$ at a redshift of 1.6 ($D_A=1700$ Mpc rad$^{-1}$). Therefore, 
if an object can be spatially resolved at this turnover redshift, it will be resolved at all higher redshifts. 

\begin{figure}
\plotone{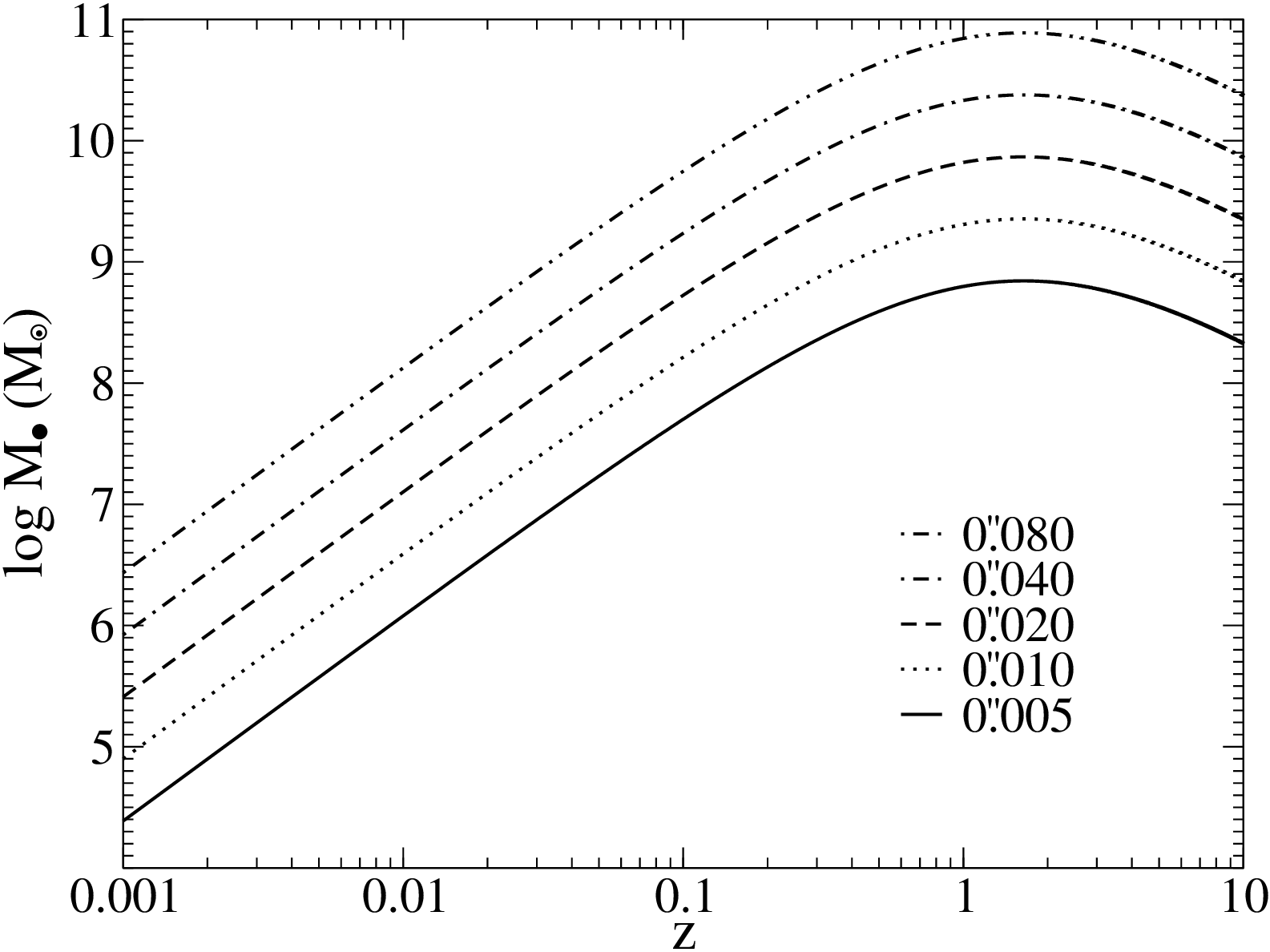}
\caption{SMBH detection requirements as a function of redshift. Each line marks the different selected spatial resolutions, given by the inset, 
needed to resolve $r_h$. $10^{8.8}M_\odot$ SMBHs can be seen by a 0\farcs005 diffraction limited telescope at all redshifts. 
}
\label{fig:2}
\end{figure}

The $M_\bullet-\sigma_\ast$ relation can be rewritten as:

\begin{equation}\label{msigma}
M_\bullet = 10^\alpha\left(\frac{\sigma_\ast}{\sigma_0}\right)^\beta
\end{equation}

where $\sigma_0 = 200\kms$ \citep{2005SSRv..116..523F}. This can be rearranged for $\sigma_\ast^2$ and substituted into Equation~\ref{rh} to show:

 \begin{equation}
D_A\theta_D = GM_\bullet\left(\frac{10^\alpha}{\sigma_0^{\beta}M_\bullet}\right)^{2/\beta}
\end{equation}

where $D_A\theta_D = r_h$. Consequently, we can solve for $M_\bullet$ and derive Equation~\ref{equ:mvz}. Therefore, for a given $\theta_D$, we can 
calculate the values of $M_\bullet$ that can be directly determined as a function of redshift. Furthermore, assuming $\alpha = 8.22$ and $\beta=4.86$ 
\citep{2005SSRv..116..523F}, Figure~\ref{fig:2} can be produced to demonstrate the resolution requirements of a direct SMBH mass estimator. It can be 
seen that $\theta_D<0\farcs01$ is required to resolve $M_\bullet=10^{9.3}M_\odot$, and $\theta_D<0\farcs005$ to resolve $M_\bullet=10^{8.8}M_\odot$, 
at all redshifts.

\begin{equation}\label{equ:mvz}
M_\bullet = \left[\frac{D_A\theta_D}{G}\left(\frac{\sigma_0^2}{10^{2\alpha/\beta}}\right)\right]^{(1-2/\beta)^{-1}}
\end{equation}

\begin{figure}
\plotone{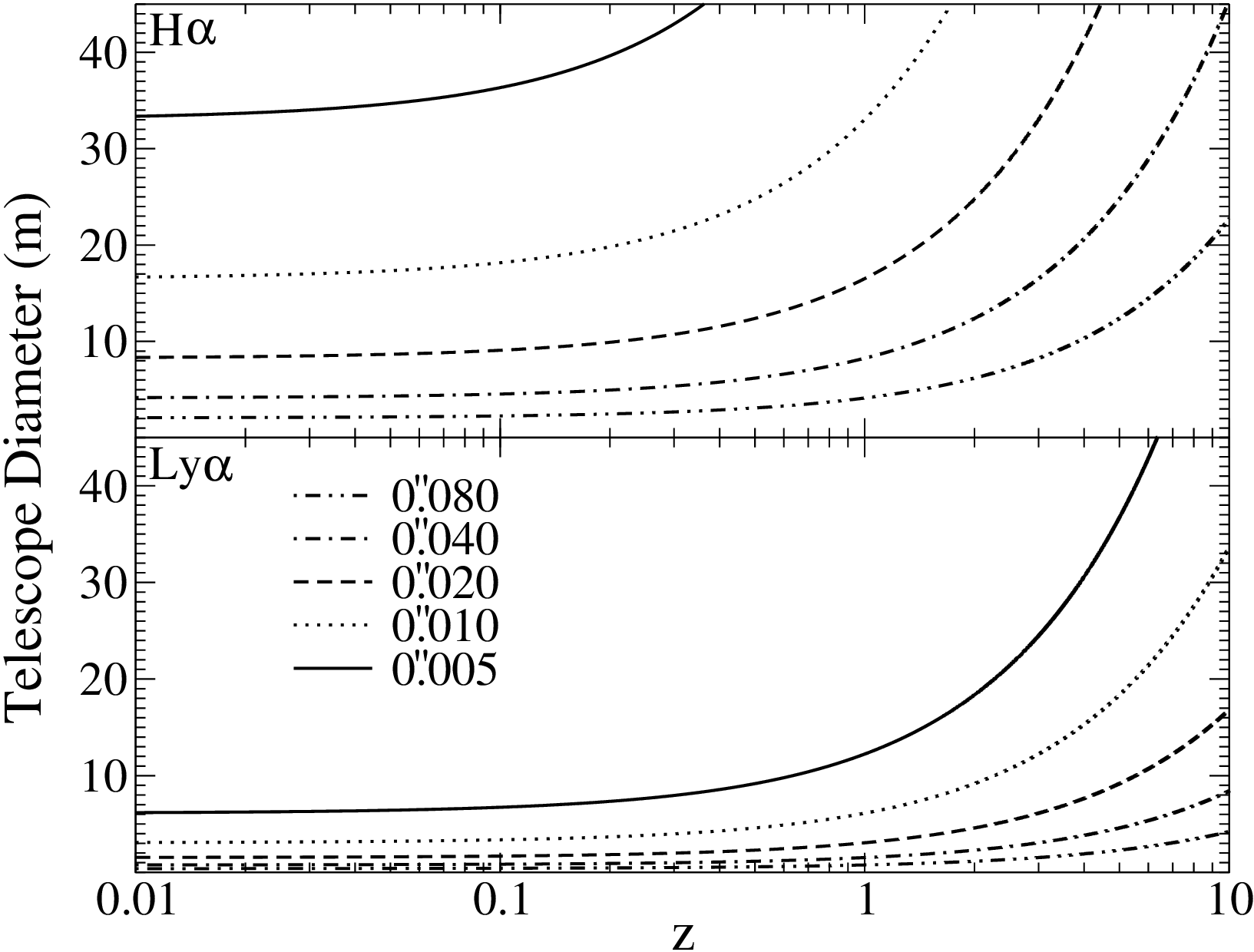}
\caption{Required telescope diameters to achieve specific resolutions as a function of redshift. {\it [Upper panel]} The redshift evolution of 
\ha\ (6563{$\rm\AA$} rest frame). {\it [Lower panel]} The redshift evolution of \lya\ (1216{$\rm\AA$} rest frame).
}
\label{fig:3}
\end{figure}

Given a specific spectral feature at a particular rest wavelength, Figure~\ref{fig:2} can be simply related the physical diameter of a telescope 
capable of observing $M_\bullet$ (at a given redshift) by using the diffraction limit equation. Figure~\ref{fig:3} demonstrates these relations 
considering the \ha\ and \lya\ emission lines that could be used for gas kinematical $M_\bullet$ estimates. These two lines are chosen to 
demonstrate the impact of being able to observe features in the UV verses the optical, i.e., ground-based verses space-based (see \S~\ref{future}). 

Figures~\ref{fig:2} and \ref{fig:3} can be used in tandem to determine what size telescope is required to observe a specific mass SMBH at a 
specific redshift using \ha\ or \lya. From Figure~\ref{fig:2} we can see that at $z=1$, a resolution of 0\farcs02 is required to detect a $10^{9.8}M_\odot$ 
HMBH. From Figure~\ref{fig:3}, observing this system using \ha\ requires a 17~m diffraction limited telescope. However, if one were to switch to 
observe \lya, then the same observations can be made using a 3.1~m primary. Alternatively, the same 17~m telescope would be able to detect 
SMBHs down to masses of $10^{8.8}M_\odot$ out past $z=1.6$ using \lya. Table~\ref{tab:1} uses this approach to present the telescope diameters 
required to observe a range of $M_\bullet$ across a range in redshift, using several different spectral features. 

\begin{deluxetable*}{lcccc|cccc|cccc|cccc}
\tablecaption{Required Telescope Diameters for SMBH Mass Estimates\label{tab:1}}
\tablewidth{0pt}
\tablehead{
\colhead{$M_\bullet$}&
\multicolumn{16}{c}{Redshift}\\
\colhead{$(M_\odot)$}&
\multicolumn{4}{c}{\lya}&
\multicolumn{4}{c}{\ha}&
\multicolumn{4}{c}{CaT}&
\multicolumn{4}{c}{HeI}\\
\colhead{}&
\colhead{0.01}&
\colhead{0.1}&
\colhead{1}&
\colhead{10}&
\colhead{0.01}&
\colhead{0.1}&
\colhead{1}&
\colhead{10}&
\colhead{0.01}&
\colhead{0.1}&
\colhead{1}&
\colhead{10}&
\colhead{0.01}&
\colhead{0.1}&
\colhead{1}&
\colhead{10}
}
\startdata
$10^5$      &      27 &  260 & 2100 &	6100 &		 140 & 1400 & 11000 & 33000 & 		 188 & 1800 & 15000 & 43000 & 		 240 & 2300 & 19000 & 54000 \\
$10^6$      &     6.9 &    68 &   540 &	1600 &		   37 &   370 &   2900 &   8500 & 		   48 &   480 &   3800 & 11000 & 	 	   61 &   600 &   4800 & 14000 \\
$10^7$      &     1.8 &    17 &   140 &	  400 &		  9.6 &     94 &     760 &   2200 & 		   13 &   122 &     980 &   2900 & 		   16 &   155 &   1200 &   3600 \\
$10^8$      &   0.46 &   4.5 &     36 &	  100 &		  2.5 &     24 &     200 &     570 & 		  3.2 &     32 &     250 &     740 & 		  4.1 &     40 &     320 &     940 \\
$10^9$      &   0.12 &   1.2 &    9.3 &	    27 &		0.64 &    6.3 &       50 &     150 & 		0.83 &    8.2 &       66 &     190 & 	 	  1.1 &     10 &       83 &     240 \\
$10^{10}$& 0.030 & 0.30 &    2.4 &	   7.0 &		0.16 &    1.6 &       13 &       38 & 		0.21 &    2.1 &       17 &       49 & 		0.27 &    2.7 &       21 &       62 \\
\enddata
\tablecomments{Telescope diameters (in meters) required to resolve specific black hole masses at specific redshifts, using specific spectral features. This assumes the telescopes are 
producing diffraction limited observations at all wavelengths, and does not take into account atmospheric absorption, i.e., a ground-based telescope cannot detect \lya\ until z$\ge$1.63.
}
\end{deluxetable*}

\section{Required Sensitivities}\label{sense}

The same cosmology that allows the angular diameter distance to turn over at $z=1.6$ also affects the observed surface brightnesses; the 
same flux is now distributed over a larger solid angle. The luminosity distance ($D_L$) is related to the angular diameter distance by 
Equation~\ref{equ:dl} \citep{1999astro.ph..5116H}. Subsequently, surface brightness drops off rapidly in a standard cosmology. 

\begin{equation}\label{equ:dl}
D_L = D_A(1+z)^2
\end{equation}

To assess how this cosmological dimming will affect $M_\bullet$ estimates we take M87 as an example. As M87 is a giant elliptical, it 
represents the class of galaxy that is expected to host the most massive SMBHs, i.e., brightest cluster galaxies \citep{2009ApJ...690..537D}. 
Therefore, based on resolving $r_h$, such giant ellipticals offer the chance to directly determine the mass of the highest redshift SMBHs. In 
addition, M87 type objects provide a challenge to both stellar dynamical models, due to low surface brightness, and gas dynamical models, 
due to weak nuclear emission lines. 

\begin{figure}
\plotone{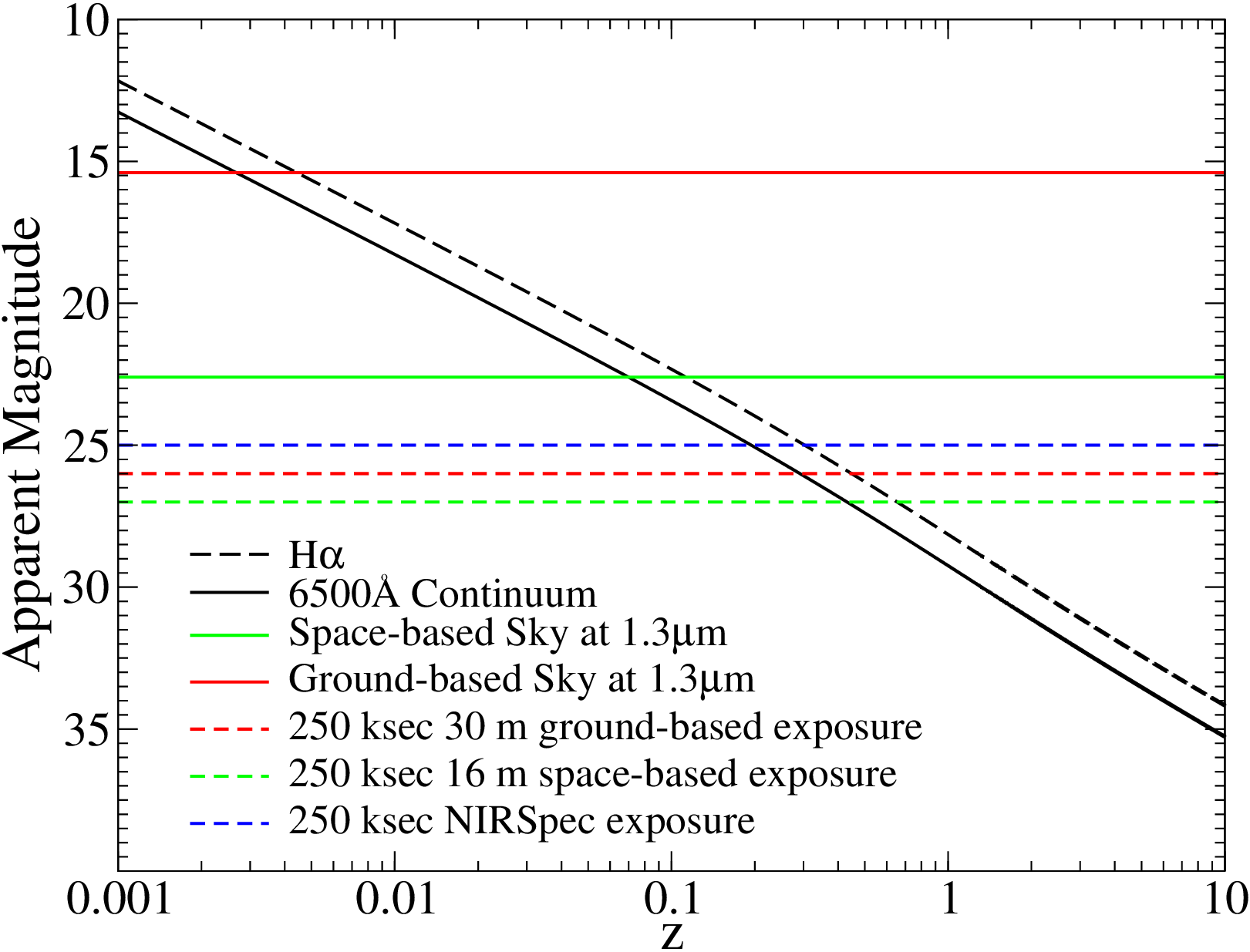}
\caption{Cosmological dimming of an M87 type object. Colored lines mark the sky backgrounds at 1.3\micron\ and the effective sensitivities of a 
30~m ground-based telescope, a 16~m space-based telescope, and JWST+NIRSpec assuming a 250~ksec exposure time. 
}
\label{fig:4}
\end{figure}

We have estimated a 6500\AA\ continuum surface brightness of 16.0mag~arcsec$^{-2}$ based on the ACS F475W and F850LP surface 
brightness profiles presented by \citet{2006ApJS..164..334F}. In addition, we have taken the \ha\ line flux of 
$0.8\times10^{-15}{\rm ergs~s^{-1}~cm^{-2}}$ as determined from STIS observations by \cite{2003ApJ...584..164S}. These estimates were 
converted to absolute values assuming a distance modulus to Virgo of 31.09 \citep{2001ApJ...546..681T}. The apparent magnitudes were 
then rescaled, assuming a constant $\nu{L_\nu}$, with luminosity distance as a function of redshift. The relation of these apparent magnitudes 
as presented, as a function of redshift, in Figure~\ref{fig:4}. 

In \S~\ref{future} we compare the abilities of both ground- and space-based telescopes to make direct SMBH mass estimates. Therefore, 
for comparison, in addition to the continuum and line fluxes for M87 as a function of redshift, we over plot the sky backgrounds as 
experienced at Mauna Kea\footnote{e.g., http://www2.gemini.edu/sciops/telescopes-and-sites/observing-condition-constraints} using the 
solid red line and at the L2 Lagrangian point using the solid green line \citep{2001AAS...19915709W}. In addition, we plot the theoretical 
1.3\micron\ sensitivities of a diffraction limited 30~m ground-based telescope (dashed red line), a 16~m space-based telescope (dashed 
green line) and JWST+NIRspec (dashed blue line). These theoretical sensitivity limits have been calculated based on achieving S/N = 10 
of a  0\farcs15 extended source in a  250~ksec exposure (M. Postman private communication). The spectral resolution is $\Re=2000$, except for 
JWST where NIRSpec ($\Re=2700$) is used. In all cases, the 16~m space-based telescope out-performs the 30~m ground-based facility, 
especially in the sky background levels. 

\begin{figure*}
\plotone{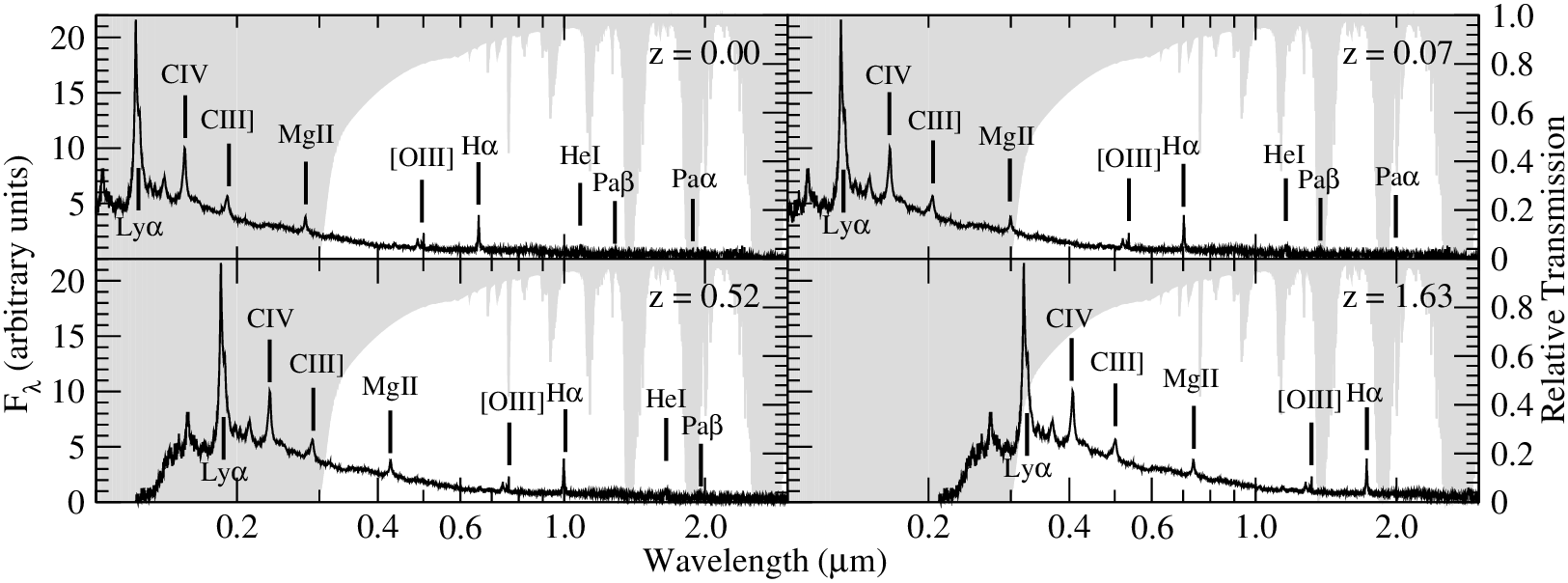}
\caption{Composite QSO spectrum at several redshifts as compared to the atmospheric relative transmission from Hawaii (grey shaded area). At  z=0.07 \ha\ becomes 
hampered by strong H$_2$0 water absorption at 0.7\micron. At z=0.52 \ha\ will be observable using NIRSpec on JWST at 1.0\micron. At z=1.62 \lya\ becomes 
visible from the ground. 
}
\label{fig:5}
\end{figure*}

\section{The Future of $M_\bullet$ Estimates}\label{future}

In \S~\ref{rh} and \S~\ref{sense} the abilities of a ground-based 30~m telescope, and a space-based 16~m telescope were briefly compared. 
The 30~m telescope represents the next generation of extremely large ground-based facilities such as the Thirty Meter Telescope (TMT),  
the 24.5~m Giant Magellan Telescope (GMT) and the 42~m European Extremely Large Telescope (E-ELT), that might be used to make 
the next step forward in our understanding of $M_\bullet$.  Each of these will provide the extremely high sensitivities required 
by stellar dynamical models. The 16~m telescope represents the possible future of UVOIR space-based observatories, the size of which 
are governed by future payload launch abilities, e.g., \citet{2009arXiv0904.0941P}.

There are some other notable upcoming facilities that will potentially provide further interesting data on SMBH masses. The Laser 
Interferometer Space Antenna (LISA, e.g., \citealt{2003AAS...202.3703H}) will be sensitive to the gravitational wave signatures of 
in-falling, and coalescing binary SMBHs. The parameters of the system (e.g., total mass) can be theoretically recovered from the periodic 
space-time strain. However, even with clearly identified signatures of binary SMBHs LISA will not be able to give details on the 
surrounding host galaxy, nor accurate enough co-ordinates for follow-up studies with other facilities.  Another interesting possibility 
is to use the Atacama Large Millimeter/Sub-millimeter Array (ALMA) to map the 2.6~mm CO kinematics. In its largest base-line 
configuration of 14.5~km, ALMA will have a spatial resolution of 0\farcs036 \citep{2008JPhCS.131a2049P}, the equivalent of a 
4.2~m diffraction limited telescope at 600~nm. The final upcoming facility that may provide a step forward in $M_\bullet$ determinations 
is the James Webb Space Telescope (JWST). In combination with NIRSpec, the 6.5~m aperture will be able to provide diffraction limited 
integral field and long-slit spectra ($\Re\sim3000$) from 1-5\micron. Therefore, \pa\ (1.88\micron) and the CO band-heads (1.5-4.7\micron) will 
be observable for possible gas and stellar dynamical models. Unfortunately, at these wavelengths the diffraction limit of JWST offers no 
spatial advantage over observations with {\it HST}. However, for 15~mag~arcsec$^{-2}$ extended sources at 8561\AA, STIS requires 
$>540$ minutes to gain a single S/N$\sim50$ spectrum. Assuming NIRSpec has a similar efficiency to STIS, the same observation could be 
made in 80 minutes, making JWST significantly more efficient for absorption line spectroscopy. This gain  in efficiency will not translate 
to the emission lines, however, due to the relative line strength of \ha\ to \pa. 

Despite the possible advances offered by ALMA and JWST, it must be noted that there have yet to be any significant attempts to 
confirm that SMBH masses derived from gas or stellar dynamics, or from multiple emission lines (e.g., \lya, \ha, \pa, [OIII]) and absorption features 
(e.g., Mgb, CaT), produce consistent results. Different spectral features will be affected by different issues in different ways. For example, in the 
case of absorption lines, AGN continua will fill shorter wavelength features, careful sky subtraction will need to be performed for longer 
wavelength features, and different features contend with template mismatching with differing levels of success. There are a few cases, 
however, in which independent, i.e., from different authors, SMBH mass estimates have been made from both stars and gas, or by using different 
spectral features (Tab.~\ref{tab:2}). Unfortunately, no significant conclusions can be drawn from such a small sample. There are a number of additional 
cases in which authors have attempted to reconcile gas and star estimates directly.  For example, in NGC~3379 \cite{2006MNRAS.370..559S} find gas and 
stellar dynamical estimates to be in agreement (although strong non-circular motions are detected in the gas disk.), and in NGC~4335 
\cite{2002AJ....124.2524V} find it difficult to match gas and stellar dynamical mass estimates. To add to the confusion, both \cite{1996ApJ...459L..57K} 
 \cite{1999MNRAS.303..495E} both use stellar models (using the same spectral features) to derived $M_\bullet=10^{9.3}$ and $M_\bullet=10^{9.0}$, 
respectively, in NGC~3115. 

\begin{deluxetable*}{lcl|clc}
\tablecaption{Independent Black Hole Mass Estimates \label{tab:2}}
\tablewidth{0pt}
\tablehead{
\colhead{Name}&
\colhead{$M_\bullet$}&
\colhead{Ref. }&
\colhead{$M_\bullet$}&
\colhead{Ref. }&
\colhead{Consistent?}
}
\startdata
NGC~3227 & $1.4^{+1.0}_{-0.6}\times10^7M_\odot$ & s-1        & $2.0^{+1.0}_{-0.5}\times10^7M_\odot$ & g-2         & Yes \\
NGC~4151 & $6.5(\pm0.7)\times10^7M_\odot$            & s-3        & $3^{+1}_{-2}\times10^7M_\odot$          & g-1         & No \\
NGC~5128 & $0.7-1.1\times10^8M_\odot$                    & g-4$^a$& $5^{+2}_{-1}\times10^7M_\odot$          & g-5$^b$ & Yes \\
\enddata
\tablecomments{Independent and peer reviewed direct black hole mass estimates using gas and stellar dynamics, or using different spectral features. 
References are ``s'' for stellar dynamics, ``g'' for gas kinematics, (a) for [SIII]$\lambda\lambda9071,9533$\AA, (b) for  H$_2$ (2.122\micron), 
(1) \cite{2006ApJ...646..754D}; 
(2) \cite{2008ApJS..174...31H}; 
(3) \cite{2007ApJ...670..105O}; 
(4) \cite{2006A&A...448..921M}; 
(5) \cite{2007ApJ...671.1329N}}
\end{deluxetable*}

While it may be possible to mitigate the effects of many of these issues, the most significant obstacle that must be faced by all ground-based 
observations remains the atmosphere. The Earth's thin protective blanket influences spatial resolution, limits spectral coverage, and affects the 
required sensitivity of ground-based instruments with transmissions that vary with wavelength. In addition, red-ward of $\sim0.7$\micron\ the sky 
begins to significantly interfere with ground-based observations due to strong OH emission lines (e.g., \citealt{1997PASP..109..614O}). The 
removal of these sky lines can introduce significant overheads to observing programs. For example, the Nod and Shuffle technique 
\citep{2001PASP..113..197G} removes sky lines with high precision, at the expense of doubling the effective exposure time.  The only way 
to over come these issues is to place an instrument in space, however,  limited spatial resolution is now being successfully addressed by 
adaptive optics (AO) systems. 

\begin{figure*}
\plotone{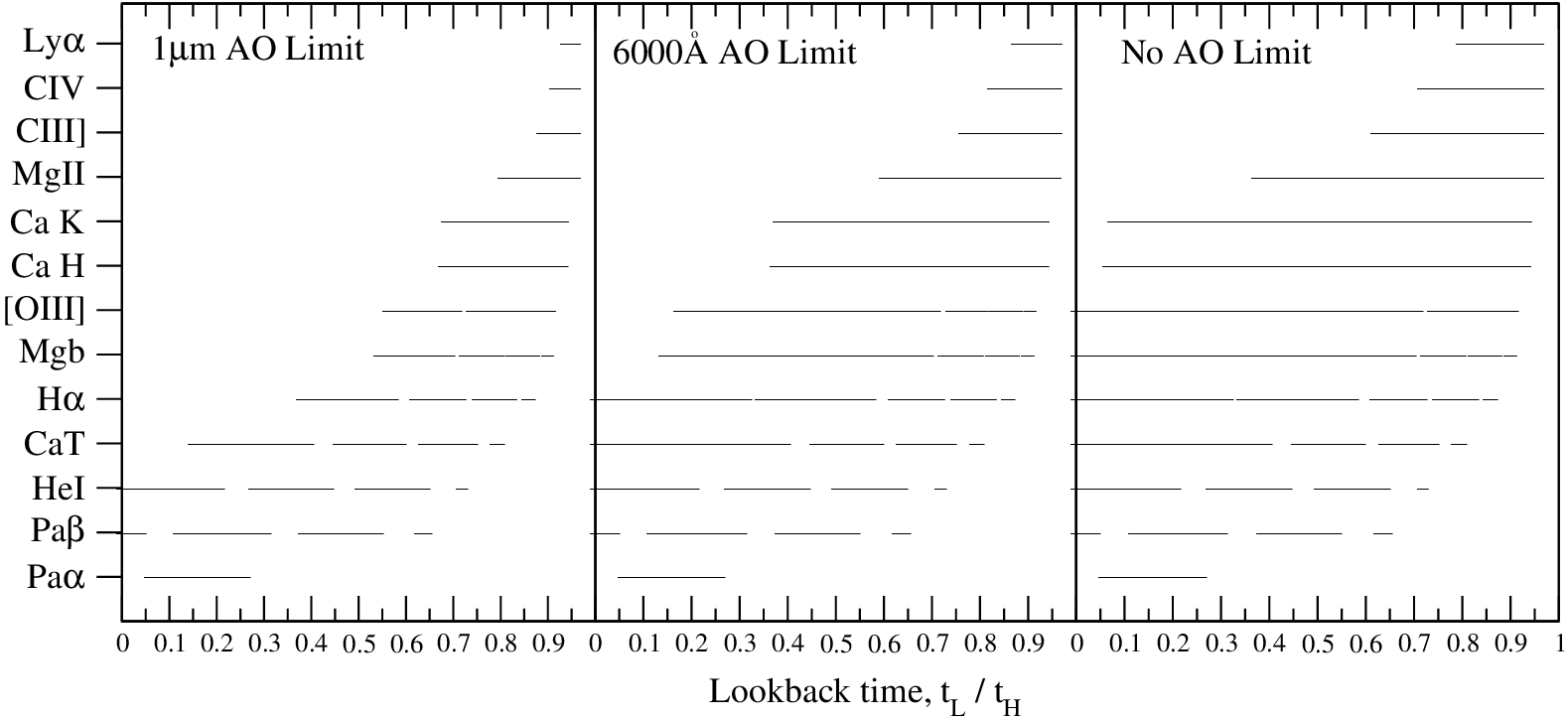}
\caption{Visibility of emission and absorption features, between 800\AA\ and 3.0\micron\ in the rest frame, that can potentially be used to 
determine $M_\bullet$, as seen by any ground-based telescope, as a function of look-back time.  The presence of a solid line indicates that the 
spectral feature is visible from the ground. Three adaptive optics dependent situations are shown. {\it [Left panel]} No advances in AO. 
{\it [Middle Panel]} Diffraction limited observations at 6000\AA. {\it [Right panel]} Diffraction limited observations all all 
wavelengths.
}
\label{fig:6}
\end{figure*}

At present AO systems are limited to the near and mid-IR, i.e., $\ge1\micron$, therefore, albeit for the significant differences in sky 
background, ground-based AO equipped facilities should be able to perform $M_\bullet$ that are comparable to JWST. However, it must be 
noted that due to guide star requirements, e.g., laser guide stars and the availability of natural guide stars for low-order tip tilt corrections, full sky 
coverage is not currently possible for AO systems. In addition, further pointing restrictions are placed on laser guide star positioning due to possible 
beam collisions between multiple systems, and due to Federal Aviation Administration and US Space Command restrictions on laser projections with 
respects to flight crew distraction and satellite interference. Finally, AO systems must be able to produce spatially and temporally stable diffraction 
limited observations to enable deep exposures over a wide field of view. 

With careful planning and advancements in future generations of AO systems, it may be possible to alleviate many of these issues and potentially 
produce diffraction limited ground-based observations at optical wavelengths. However, regardless of the future abilities of AO systems, 
ground-based observations will always be limited by atmospheric transmission. In order to demonstrated these limitations, Figure~\ref{fig:5} shows 
a composite QSO spectrum as compared to an atmospheric transmission model. The QSO spectrum combines SDSS data \citep{2001AJ....122..549V} 
with the near-infrared (NIR) data of \cite{2006ApJ...640..579G}. The \cite{2006ApJ...640..579G} data has been reduced by 0.52 flux units and chopped at 8556\AA\ to 
make a smooth transition with the SDSS data. The emission lines that could be used by gas kinematics models are marked, although we note that the 
host bulge absorption features that trace the stellar dynamics at $\sim$3950\AA\ (Ca H\&K), $\sim$5175\AA (Mgb), $\sim$8570\AA\ (CaT) and from 
1.5-4.7\micron\ (CO bands) will not be seen on the scale used. The atmospheric model was generated using MODTRAN \citep{1999SPIE.3756..348B} 
assuming conditions typically found at observatory altitudes on Hawaii. 

The QSO spectrum is shown at z=0.00 and z=0.07, when \ha\ becomes hampered by strong H$_2$0 water absorption at 0.7\micron. In addition, 
z=0.52 is shown as \ha\ will be observable using NIRSpec on JWST at 1.0\micron. \pa\ is observable by NIRSpec at z=0 but the diffraction limit of 
JWST at 1.88\micron\ is 0\farcs07, the same as {\it HST} at \ha. Figure~\ref{fig:5} also shows the QSO spectrum at z=1.63, when \lya\ starts to 
suffer less than 50\% transmission loss at 3200\AA. 

To more precisely determine which spectral features are observable from the ground, and at what look-back time, in Figure~\ref{fig:6} we have simply 
multiplied the QSO spectrum by the relative atmospheric transmission between 800\AA\ and 3.0\micron. If a solid line is present, then the indicated 
spectral feature is observable at the indicated look-back time. 

In comparing a 16~m to a 30~m telescope we have considered the difference in limiting magnitudes between the two, according to:

\begin{equation}\label{equ:mlim}
\Delta{m} = 5\log\left(\frac{d_1}{d_2}\right)
\end{equation}

where $d_1$=30~m and $d_2$=16~m, and therefore $\Delta{m}$=1.37 magnitudes \citep{2003aste.book.....K}. The value of $\Delta{m}$ is the 
equivalent of knowing the change in magnitude as a result of atmospheric extinction. It is simple to show that $\Delta{m} = k(\lambda)\sec{Z}$, where  
$k(\lambda)$ is the wavelength dependent extinction coefficient, and $Z$ is zenith distance.  In this case, we know that the zenith distance is zero 
and that  $k(\lambda) = 1$ as it is taken into account by the relative atmospheric transmission model. Therefore, we can see that the inverse of 
$\Delta{m}$ is equal to the relative atmospheric transmission (0.73) that a ground-based 30~m can suffer from before it's aperture advantage is lost 
to a space-based 16~m telescope. Consequently, in Figure~\ref{fig:6}, if the atmospheric transmission is less than 0.73 then the spectral feature is 
considered blocked from ground-based observations. 

Figure~\ref{fig:6} shows three different cases for the future of AO systems. In the first case, there is no significant advance in AO technology, and 
observations are only considered to be diffraction limited at 1.0\micron. In the second case, AO systems are assumed to be able to produce diffraction 
limited observations down to 6000\AA. In the final case AO systems are unlimited and produce diffraction limited observations at all wavelengths. 

Look-back time (e.g., \citealt{1999astro.ph..5116H}) is plotted in Figure~\ref{fig:6}, assuming $\Omega_k=0.00$,  as it more directly shows the cosmic 
era over which $M_\bullet$ can be measured. As demonstrated by Equation~\ref{equ:lookback}, the look-back time, $t_L$ is similar to 
Equation~\ref{equ:ztoda} and proportional to the Hubble time, $t_H$. 

\begin{equation}\label{equ:lookback}
t_L = t_H \int^z_0 \frac{dz'}{(1+z')\sqrt{\Omega_M(1+z')^3 + \Omega_\lambda}}
\end{equation}

Figure~\ref{fig:6} is structured so that the shortest wavelength spectral features, and therefore the features that offer the greatest spatial resolution 
per aperture, are at the top. In addition, the closest, and therefore most well resolved, SMBHs will be found on the left hand side of each panel., 
For reference, the $D_A$ turnover at z=1.6 (Fig.~\ref{fig:1}) is at a look-back time of $t_L$=0.7$t_H$. In all AO cases, the top left 
areas of each panel in Figure~\ref{fig:6} (that can be filled by space-based observations and offer the greatest spatial resolution) remain empty. 

\begin{figure}
\plotone{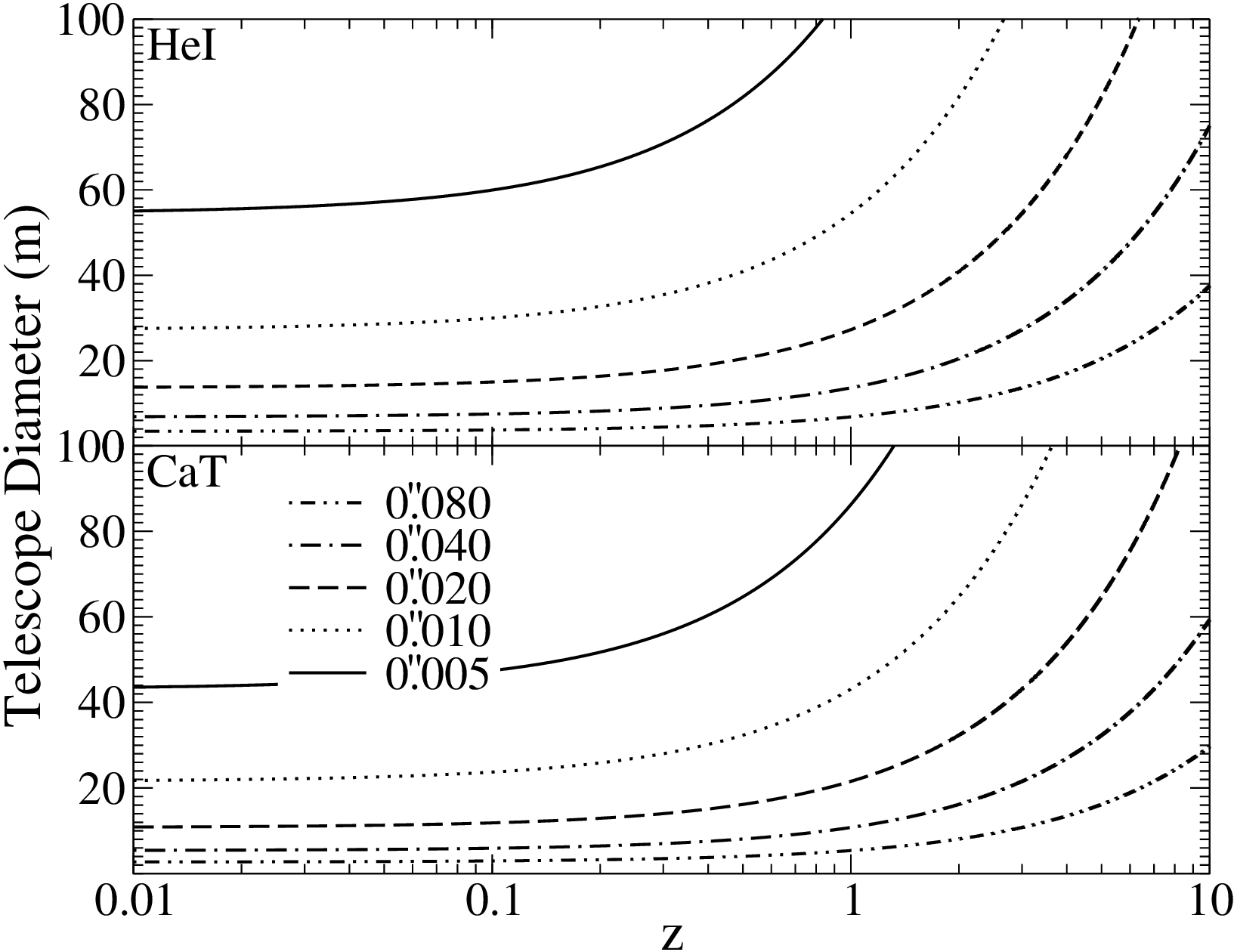}
\caption{Required telescope diameters to achieve specific resolutions as a function of redshift. {\it [Upper panel]} The redshift evolution of 
HeI (1.1\micron\ rest frame). {\it [Lower panel]} The redshift evolution of CaT ($\sim$8567\AA\ rest frame).
}
\label{fig:7}
\end{figure}

However, Figure~\ref{fig:6} demonstrates that multiple spectral features can theoretically be used to determine $M_\bullet$ across a significant 
cosmic era. With unlimited AO abilities, \ha\ and Mgb can be used to make gas and stellar dynamical models across $\sim90\%$ of cosmic history. 
In addition,  \ha\ can be used to cover the same era if AO systems can provide diffraction limited observations at 6000\AA. Stellar dynamical 
models will, with 6000\AA\ diffraction limited observations, still be able to cover this era by using a combination of Mgb and CaT estimates. Using the 
current AO abilities of 1.0\micron, then ground-based observations will need to use a combination of HeI, CaT, Mgb and \ha\ to cover a significant portion 
of cosmic history. In Figure~\ref{fig:7} we demonstrate the size of aperture required to achieve these observations, for HeI and CaT, using the same 
approach as for Figure~\ref{fig:3}. 

For comparison with Figure~\ref{fig:3}, i.e., using \ha\ and \lya, Figures~\ref{fig:2} and \ref{fig:7} can now be used together in order to determine what 
size telescope is required to observe a specific mass SMBH, at a specific redshift, using HeI or CaT. As seen in Figure~\ref{fig:3}, switching between 
spectral features dramatically affects SMBH detection efficiency. For example, to make the same observations as a 16~m, i.e., \ha\ at z=1 with a spatial 
resolution of 0\farcs02 and a mass of $10^{9.8}M_\odot$, a 27~m diffraction limited telescope is needed using HeI. Switching to observe CaT then requires 
a 21~m primary. The same observations for \pa\ (which has a lower line flux than HeI) require a 47~m diffraction limited telescope.  Therefore, a ground-based 
30~m telescope using existing AO abilities, will be able to theoretically determine the masses of high mass SMBHs and HMBHs out to high redshifts using 
HeI. However, as seen in Figure~\ref{fig:5}, the line strength of HeI is significantly weaker than higher energy emission lines. HeI will therefore more rapidly 
succumb to the background limits shown in Figure~\ref{fig:4}.

Table~\ref{tab:3} summarizes the SMBH detection abilities of several space-based telescopes and a 30~m ground-based telescope that can achieve 
diffraction limited observations from 3200\AA\ (where \lya\ suffers from less than 50\% atmospheric transmission loss) and 6000\AA, to 1.3\micron, 
where atmospheric transmission again impedes continuous redshift coverage. For the minimum detectable SMBH masses, a distance of 100Mpc is chosen 
in order to cover the Coma cluster, where a large range of SMBHs are likely to reside. Considering only the gas kinematics technique, the one most 
suited to high redshift $M_\bullet$ estimates due to emission line strengths, it can be seen that models using the strong hydrogen recombination lines are best 
supplied data by space-based observatories. Such facilities provide continuous redshift coverage using \lya\ and can therefore observe significantly 
smaller SMBHs out to significantly higher redshifts. Due to atmospheric transmission, at the distance of the Coma cluster, the 30~m ground-based 
telescope can only reach down to SMBHs of $10^{6.7}M_\odot$ compared to the $10^{6.0}M_\odot$ of the 16~m.

\begin{deluxetable}{lcccccc}
\tablecaption{SMBH Detection Abilities \label{tab:3}}
\tablewidth{0pt}
\tablehead{
\colhead{Telescope}&
\multicolumn{3}{c}{Redshift range}&
\multicolumn{3}{c}{Min. $\log{M_\bullet}$}\\
\colhead{}&
\colhead{\lya}&
\colhead{\ha}&
\colhead{HeI}&
\colhead{\lya}&
\colhead{\ha}&
\colhead{HeI}
}
\startdata
{\it HST} & 0.0--7.5 & 0.0--0.6 & \nodata  & 7.0        & 8.6 & \nodata \\
JWST           & \nodata  & 0.5--6.6 & 0.0--2.8 & \nodata & 8.5 & 8.5     \\
16SB            & 0.0--7.5 & 0.0--0.6 & \nodata  & 6.0        & 7.1 & \nodata \\
TMT(a)       & 1.6--4.8 & 0.0--0.1 & 0.0-0.2  & 8.4        & 6.7 & 7.1 \\
TMT(b)       & 3.9--4.8 & 0.0--0.1 & 0.0-0.2  & 8.8        & 6.7 & 7.1 \\
\enddata
\tablecomments{SMBH detection abilities, including redshift coverage as a function of prominent emission lines, for current and pending telescopes. 
16SB: 16~m space-based telescope. TMT: 30~m ground-based telescope, diffraction limited at (a) 3200\AA\ and (b) 
6000\AA. ``$z$ range'': the continuous observable redshifts for {\it HST}+STIS (1150--10,300\AA), and JWST+NIRSpec (1.0--5.0\micron). 
We give 16SB an instrument equivalent to STIS, and limit TMT to 3200\AA\ -- 1.3\micron. ``Min. $\log{M_\bullet}$'': the minimum $M_\bullet$ 
detectable at 100Mpc. 
}
\end{deluxetable}

In addition to the data in Table~\ref{tab:3}, we can see that by using \ha\ the $\theta_D$ of {\it HST} can detect $10^{10}M_\odot$ HMBHs 
out to a maximum distance of z=0.15. However, a sensitivity of $\sim25$~mag~arcsec$^{-2}$ is required to detect an object similar to M87 at 
this distance (Fig.~\ref{fig:4}). Therefore, {\it HST} can only determine $M_\bullet$ in the brightest most massive QSOs at this distance, of 
which there are few. Using \lya, {\it HST}'s distance limit changes to $10^{9.7}M_\odot$ SMBHs at z=0.6. JWST will not be able to observe 
\ha\ until z=0.6, but will be able to observe HeI, \pb\ and \pa\ across a large redshift range. The longer wavelengths of these features, however, 
negates the increase in the JWST aperture as compared to {\it HST}, and results in no gain in spatial resolution. The JWST advantage will 
come from the greater collecting area. Naturally, a 16~m space-based mission would not be as limited as {\it HST} and JWST. It would be able 
to detect $10^{9.8}M_\odot$ SMBHs to $z=1.0$ using \ha, and $10^{9.4}M_\odot$ SMBHs to $z=10$ using \lya. It would also allow a search 
for IMBHs in the local group out to 4.5 Mpc. 

Irrespective of the SMBH detection abilities of space- verses ground-based telescopes, it is clear that the consistency of $M_\bullet$ modeling 
techniques, using multiple spectral features, is vital for continued investigations into the role of SMBHs in galaxy formation and evolution. For example, 
for the abilities listed in Table~\ref{tab:3} to be reached, it must first be shown that both \lya\ and HeI can be used to determine $M_\bullet$. In addition, 
as gas disks are not ubiquitous at the nuclei of galactic bulges, gas kinematical models must also be reconciled with stellar dynamical models using 
at least both Mgb and CaT. The next step in $M_\bullet$ investigations must therefore be to produce consistent SMBH masses, from different methods 
and spectral features, to pave the way for more complete SMBH investigations using both space- and ground-based facilities. Space is {\it required} 
for \lya\ $M_\bullet$estimates, but ground-based AO assisted observations of HeI and the CO band-heads will be useful, provided the aperture is large enough 
to compensate for the longer wavelengths and lower line fluxes. Due to the large wavelength coverage (1150--10,300\AA) {\it HST}+STIS provides an 
excellent opportunity to complete a $M_\bullet$ calibration for \lya\ and both Mgb and CaT. In addition, the calibration of NIR lines would be an excellent 
program for JWST+NIRspec. 

\section{Conclusions}\label{conclusions}

Our understanding of the co-evolution of SMBHs and galactic bulges has been driven by the discovery of scaling relations between the two. However, there 
are a number of fundamental concerns with these relations that will have an impact on galaxy formation and evolution scenarios. The future of direct $M_\bullet$ 
determinations is therefore likely to be dominated by five key questions. Are IMBHs and HMBHs consistent with the bulge scaling relations defined 
by SMBHs? Are bulge scaling relations linear? Have bulge scaling relations evolved to their present form? Are there upper and lower limits to the BHMF? 
What are the direct $M_\bullet$ estimates in QSOs, and do they follow the bulge scaling relations? It is therefore necessary to expand our abilities to 
determine the masses of a range of SMBHs across as wide a cosmic history as possible. 

We have considered the requirements for an effective SMBH detector that uses current gas and stellar dynamical models and techniques to derive 
$M_\bullet$. Cosmological models have been coupled with the sphere of influence argument to determine the aperture sizes needed in order to 
resolve SMBHs of specific masses at a range of redshifts. In addition, we have considered the sensitivities that such facilities will need. 

It has been demonstrated that the limits of $M_\bullet$ estimates from {\it HST} are being reached and cannot be significantly expanded by 
JWST. Therefore, additional facilities are needed to directly determine whether IMBHs, HMBHs and QSOs follow the same scaling relations 
as typical galactic bulges, and whether locally established scaling relations are linear and cosmically constant. 
 
A 30~m ground-based optical-NIR observatory (e.g., the forthcoming TMT) has been directly compared to a UVOIR 16~m space-based observatory. 
For either facility to be an effective SMBH detector, current modeling techniques must be directly compared with each other to ensure consistent mass 
estimates from different models using different spectral features. 

If consistency can be established, then a space-based 16~m telescope can theoretically be used to estimate $M_\bullet$ using \lya\ out to z=10, or across 96\% of cosmic 
history (depending on the instruments, and limited to $M_\bullet\ga10^{9.4}M_\odot$). This facility will also be able to detect  $M_\bullet\sim10^{6.0}M_\odot$ 
at a distance of the Coma Cluster, and  $M_\bullet\sim10^{4.5}M_\odot$ in the Local Group. A 30~m ground-based telescope will need to use multiple spectral 
features in order to cover the same amount of cosmic history, even in the face of significant advances in AO systems. Due to limitations imposed by 
atmospheric transmission, the 30~m can theoretically be used to estimate $M_\bullet$ using \lya\ from z=1.6 to z=4.8, or across 20\% of cosmic 
history (depending on the instruments). For increased cosmic history coverage, longer wavelength spectral features will need to be used at the cost of SMBH 
detection efficiency. The 30~m will also be able to detect  $M_\bullet\sim10^{6.7}M_\odot$ at a distance of the Coma Cluster, and  $M_\bullet\sim10^{4.2}M_\odot$) 
in the Local Group. 

However, the abilities of the ground-based 30~m are critically dependent on future advances in AO systems. As AO abilities increase toward the NIR, so 
does the effective spatial resolution, and therefore the SMBH detection efficiency. Considering current AO systems, the 30~m is limited to observing 
HeI. In this case, the 30~m will be able to detect  $M_\bullet\sim10^{7.1}M_\odot$ at a distance of the Coma Cluster. In addition, emission lines at these 
wavelengths are significantly weaker than features blue-ward of \ha. As a consequence, the advantages of the large aperture are lost as it must integrate for 
longer to gain the same S/N that a 16~m would have observing \lya. The ultimate advantage for a space-based telescope, regardless of AO systems, 
then becomes the limitations to sensitivities as determined by sky backgrounds. As shown in Figure~\ref{fig:4}, the magnitude limits for an object such 
as M87 are reached very quickly from the ground. It will then become incredibly challenging, considering all the potential overheads (such as nod and shuffle), 
to gain the required S/N for dynamical models in a reasonable amount of observing time.

\acknowledgments

We thank Marc Postman for providing the telescope theoretical sensitivities, and for useful discussions and suggestions. We thank Rolando Raque\~{n}o of the 
Digital Imaging and Remote Sensing Lab, Center for Imaging Science, Rochester Institute of Technology, for the MODTRAN atmospheric transmission model. 
Finally, we thank David Axon, Alessandro Marconi, David Merritt, Zoran Ninkov, and the referee for useful discussions, comments and suggestions. Support for 
this work was provided by proposal number HST-AR-10935.01 awarded by NASA through a grant from the Space Telescope Science Institute, which is operated by 
the Association of Universities for Research in Astronomy, Incorporated, under NASA contract NAS5-26555.  

\bibliographystyle{apj}
\bibliography{batcheldor}

\end{document}